\documentclass[aps,prc,twocolumn,showpacs,floatfix,nofootinbib,preprintnumbers,superscriptaddress,amsmath,amssymb,longbibliography]{revtex4-1}

\usepackage{graphicx}
\usepackage{epsfig}
\usepackage{bm}
\usepackage{color}
\usepackage{float}
\usepackage{dcolumn}
\usepackage{multirow} 
\usepackage{hyperref}
\usepackage{nicefrac}

\graphicspath{{.}{./figs/}}

\newcommand{\be}{\begin{equation}}
\newcommand{\ee}{\end{equation}}
\newcommand{\ba}{\begin{eqnarray}}
\newcommand{\ea}{\end{eqnarray}}

\begin{document}


\title{Effective field theory in the harmonic oscillator basis}


\author{S.~Binder} \affiliation{Department of Physics and
   Astronomy, University of Tennessee, Knoxville, TN 37996, USA}
 \affiliation{Physics Division, Oak Ridge National Laboratory, Oak
   Ridge, TN 37831, USA}

\author{A.~Ekstr\"om} \affiliation{Department of Physics and
   Astronomy, University of Tennessee, Knoxville, TN 37996, USA}
 \affiliation{Physics Division, Oak Ridge National Laboratory, Oak
   Ridge, TN 37831, USA}

 \author{G.~Hagen} \affiliation{Physics Division, Oak Ridge National
   Laboratory, Oak Ridge, TN 37831, USA} \affiliation{Department of
   Physics and Astronomy, University of Tennessee, Knoxville, TN
   37996, USA}

 \author{T.~Papenbrock} \affiliation{Department of Physics and
   Astronomy, University of Tennessee, Knoxville, TN 37996, USA}
 \affiliation{Physics Division, Oak Ridge National Laboratory, Oak
   Ridge, TN 37831, USA}

 \author{K.~A.~Wendt} \affiliation{Department of Physics and
   Astronomy, University of Tennessee, Knoxville, TN 37996, USA}
 \affiliation{Physics Division, Oak Ridge National Laboratory, Oak
   Ridge, TN 37831, USA}


\begin{abstract}
 We develop interactions from chiral effective field theory (EFT) that
 are tailored to the harmonic oscillator basis.  As a consequence,
 ultraviolet convergence with respect to the model space is
 implemented by construction and infrared convergence can be achieved
 by enlarging the model space for the kinetic energy.
 In oscillator EFT, matrix elements of EFTs formulated for continuous
 momenta are evaluated at the discrete momenta that stem from the
 diagonalization of the kinetic energy in the finite oscillator space.
 By fitting to realistic phase shifts and deuteron data we construct
 an effective interaction from chiral EFT at next-to-leading order.
 Many-body coupled-cluster calculations of nuclei up to $^{132}$Sn
 converge fast for the ground-state energies and radii in
 feasible model spaces.
\end{abstract}


\maketitle


\section{Introduction}

The harmonic oscillator basis is advantageous in nuclear-structure
theory because it retains all symmetries of the atomic nucleus and
provides an approximate mean-field related to the nuclear shell
model. 
However, interactions from chiral effective field theory
(EFT)~\cite{epelbaum2009,machleidt2011} are typically formulated in
momentum space, while the oscillator basis treats momenta and
coordinates on an equal footing, thereby mixing long- and short-ranged
physics. This incommensurability between the two bases is not only of
academic concern but also makes oscillator-based {\it ab initio}
calculations numerically expensive.
Indeed, the oscillator basis must be large enough
to accommodate the nucleus in position space as well
as to contain the high-momentum contributions
of the employed interaction.
Furthermore, one needs to perform computations at different values of
the oscillator spacing $\hbar\omega$ to gauge model-space
independence of the computed
results~\cite{maris2009,hagen2010b,jurgenson2013,roth2014}.
Several methods have been proposed to alleviate these problems.
Renormalization group transformations, for instance, are routinely
used to ``soften'' interactions~\cite{bogner2003,bogner2007,roth2014}, and many
insights have been gained through these
transformations~\cite{jurgenson2009}. However, such transformations of
the Hamiltonian and observables~\cite{lisetskiy2009,schuster2014} add
one layer of complexity to computations of nuclei.

One can contrast the effort of computations in the
oscillator basis to, for instance, computations in nuclear lattice
EFT~\cite{lee2009}.  Here, the effective interaction is tailored to
the lattice spacing and, thus, to the ultraviolet (UV) cutoff, and well-known
extrapolation formulas~\cite{luscher1985} can be used to estimate
corrections due to finite lattice sizes. 
The lattice spacing is fixed once, reducing the computational
expenses. This motivates us to seek a similarly efficient approach for
the oscillator basis, i.e., to formulate an EFT for nuclear
interactions directly in the oscillator basis.

In recent years, realistic {\it ab initio} nuclear computations pushed
the frontier from light $p$-shell
nuclei~\cite{navratil2009,barrett2013} to the medium-mass
regime~\cite{hagen2012b,holt2012,wienholtz2013,soma2013b,lahde2014,hagen2014,hergert2014,hagen2015}. At
present, the precision of computational methods considerably exceeds
the accuracy of available interactions~\cite{binder2013b}, and this is
the main limitation in pushing the frontier of {\it ab initio}
computations to heavy nuclei. To address this situation, several
efforts, ranging from new optimization protocols for chiral
interactions~\cite{ekstrom2013,ekstrom2015,carlsson2015} to the
inclusion of higher orders~\cite{entem2015,epelbaum2015} to the
development of interactions with novel regulators~\cite{gezerlis2014}
are under way.
To facilitate the computation of heavy nuclei we propose to
tailor interactions from chiral EFT to the oscillator basis.

There exist several proposals to formulate EFTs in the oscillator
basis.  Haxton and coworkers proposed the oscillator based effective
theory (HOBET)~\cite{haxton2000,haxton2002,
  haxton2007,haxton2008}. They focused on decoupling low- and
high-energy modes in the oscillator basis via the Bloch-Horowitz
formalism, and on the resummation of the kinetic energy to improve the
asymptotics of bound-state wave functions in configuration space.  The
HOBET interaction is based on a contact-gradient expansion, and the
matrix elements are computed in the oscillator basis.  The resulting
interaction exhibits a weak energy dependence.  The Arizona
group~\cite{stetcu2007,stetcu2010,rotureau2012} posed and studied
questions related to the UV and infrared (IR) cutoffs imposed by the
oscillator basis, and developed a pion-less EFT in the oscillator
basis. This EFT was also applied to harmonically trapped
atoms~\cite{rotureau2010}. In this approach, the interaction matrix
elements are also based on a contact-gradient expansion and computed
in the oscillator basis. Running coupling constants depend on the UV
cutoff of the employed oscillator basis.  In a sequence of other
papers, T\"olle {\it et al.} studied harmonically trapped few-boson
systems in an effective field theory based on contact interactions
with running coupling constants~\cite{tolle2011,tolle2013}.
 
Our oscillator EFT differs from these approaches.  For the interaction
we choose an oscillator space with a fixed oscillator frequency
$\omega$ and a fixed maximum energy $(N_{\rm max} + \nicefrac{3}{2})
\, \hbar\omega$.  The matrix elements of the interaction are taken
from an EFT formulated in momentum space and evaluated at the discrete
momentum eigenvalues of the kinetic energy in this fixed oscillator
space.  This reformulation, or projection, of a momentum-space EFT
onto a finite oscillator model space requires us to re-fit the
low-energy coefficients (LECs) of interactions at a given order of the
EFT. We determine these by an optimization to scattering phase shifts
(computed in the finite oscillator basis via the $J$-matrix
approach~\cite{heller1974,shirokov2004}) and from deuteron properties.
The power counting of the oscillator EFT is based on that of the
underlying momentum-space EFT. The finite oscillator space introduces
IR and UV
cutoffs~\cite{stetcu2007,coon2012,furnstahl2012,more2013,furnstahl2014,koenig2014,furnstahl2015,wendt2015},
and these are thus fixed for the interaction. 

In practical many-body calculations we will keep the oscillator frequency and
the interaction fixed at $\hbar\omega$ and $N_{\rm max}$, 
but employ the kinetic energy in larger model spaces.
This increase of the model space increases (decreases) its UV (IR)
cutoff but does not change any interaction matrix elements 
and thus leaves the IR and UV cutoff of the interaction unchanged.
  As UV convergence of the many-body calculations depends on the matrix
elements of an interaction~\cite{koenig2014}, oscillator EFT
guarantees this UV convergence by construction
because no
  new potential matrix elements
enter beyond $N_{\rm max}$. 
We stress that our notion of UV convergence relates to the
convergence of the many-body calculations and
 should not be confused with the expectation that
  observables are independent of the regulator or cutoff.  Infrared
convergence builds up the exponential tail of bound-state wave
functions in position space, as the effective IR cutoff of a finite
nucleus is set by its radius.
Thus, the increase of the model space
for the kinetic energy achieves IR convergence. Regarding IR
convergence, oscillator EFT is similar to the \mbox{HOBET} of
Ref.~\cite{haxton2007}. In practice, IR-converged values for
bound-state energies and radii can be obtained applying
``L\"uscher-like'' formulas for the oscillator basis
\cite{furnstahl2012}.

We view oscillator EFT similar to lattice EFT~\cite{lee2009}. The
latter constructs an interaction on a lattice in position space while
the former builds an interaction on a discrete (but non-equidistant)
mesh in momentum space.  In both EFTs the UV cutoff of the interaction
is fixed once and for all, and LECs are adjusted to scattering data and
bound states. The increase in lattice sites achieves IR convergence in
lattice EFT, while the increase of the number of oscillator shells
achieves IR convergence in oscillator EFT.

As we will see, the resulting EFT interaction in the oscillator basis
exhibits a fast convergence, similar to the phenomenological JISP
interaction~\cite{shirokov2007}.
 From a practical point of view, our
approach to oscillator EFT allows us to employ all of the existing
infrastructure developed for nuclear calculations.

The discrete basis employed in this paper is actually 
the basis set of 
a discrete
variable representation
(DVR)~\cite{harris1965,dickinson1968,light1985,baye1986,littlejohn2002,light2007,bulgac2013}
in momentum space. 
While coordinate-space DVRs are particularly useful
and popular in combination with local potentials, the results of this
paper suggest that DVRs are also useful in 
momentum-space-based EFTs, because they facilitate the evaluation
of matrix elements.

This paper is organized as follows. In Section~\ref{theo} we analyze
the momentum-space structure of a finite oscillator basis.  In
Section~\ref{optimize} we validate our approach by reproducing an
interaction from chiral EFT at next-to-leading order (NLO).  In
Section~\ref{results} we construct a NLO interaction from realistic
phase shifts and employ this interaction in many-body calculations,
demonstrating that converged binding energies and radii can be
obtained for nuclei in the mass-100 region  
in model spaces with $N_{\rm max} =$~10 to 14 without any
 further renormalization. We
finally present our summary in Section~\ref{summary}.


\section{Theoretical considerations}
\label{theo}

In this Section we present the theoretical foundation of an EFT in
the oscillator basis. We derive analytical expressions for the momentum
eigenstates and eigenvalues in finite oscillator spaces and present useful
formulas for interaction matrix elements.

\subsection{Momentum states in finite oscillator spaces}
The radial wave functions $\langle r,l|n,l\rangle=\psi_{nl}(r)$ of
oscillator basis states $|n,l\rangle$ can be represented in terms of
generalized Laguerre polynomials $L_n^{l+\nicefrac{1}{2}}$ as
\ba
\label{waveR}
\lefteqn{\psi_{n,l} (r) =}\nonumber\\
&& (-1)^n\sqrt{2n!\over \Gamma(n+l+\nicefrac{3}{2})b^3} \left({r\over b}\right)^l e^{-{1\over 2}\left({r\over b}\right)^2} 
L_n^{l+\nicefrac{1}{2}}\left(\tfrac{r^2}{ b^2}\right) .
\ea
Here, $b\equiv \sqrt{\hbar/(m\omega)}$ is the oscillator length
expressed in terms of the nucleon mass $m$ and oscillator frequency
$\omega$. In what follows, we use units in which $\hbar=1$.

In free space, the spherical eigenstates of the momentum operator $\hat{p}$ are
denoted as $|k,l\rangle$ with $k$ being continuous. The corresponding wave
functions 
\be
\label{bessel}
\langle r,l|k,l\rangle = \sqrt{2\over \pi}j_l(kr) 
\ee
are spherical Bessel functions $j_l$ up to a normalization factor.
These continuum states are normalized as
\be
\langle k',l|k,l\rangle = \int\limits_0^\infty {\rm d}r r^2 \langle k',l|r,l\rangle \langle r,l|k,l\rangle = 
{\delta(k-k')\over kk'} .
\ee
Introducing the momentum-space representation of the radial
oscillator wave functions via the Fourier-Bessel transform
\ba
\lefteqn{\tilde{\psi}_{n,l}(k) \equiv \int\limits_0^\infty {\rm d} r r^2 \langle k,l|r,l\rangle \psi_{n,l}(r)} \\
&&=\sqrt{2n! b^3\over \Gamma(n+l+\nicefrac{3}{2})} \left(kb\right)^l e^{-{1\over 2}k^2b^2} L_n^{l+\nicefrac{1}{2}}\left(k^2b^2\right) \nonumber,
\ea
enables us to expand the continuous momentum states~(\ref{bessel}) in
terms of the oscillator wave functions
\be
\label{bessel2}
\langle r,l|k,l\rangle = \sum_{n=0}^\infty
\tilde{\psi}_{n,l}(k)\psi_{n,l}(r) . 
\ee
As we want to develop an EFT, it is most important to understand the
squared momentum operator $\hat{p}^2$. An immediate consequence of
Eq.~(\ref{bessel}) is of course that the spherical Bessel functions
are also eigenfunctions of $\hat{p}^2$ with corresponding eigenvalues $k^2$, 
i.e.,  \mbox{$\hat{p}^2j_l(kr)=k^2j_l(kr)$}.  
In the oscillator basis, the matrix representation of
$\hat{p}^2$ is tri-diagonal, with elements
\ba
\label{p2mat}
\langle n',l |\hat{p}^2| n,l\rangle &=& b^{-2} \bigg[
(2n+l+\nicefrac{3}{2})\delta_n^{n'} \nonumber\\
&&-\sqrt{n(n+l+\nicefrac{1}{2})}\delta_n^{n'+1}\nonumber\\
&&-\sqrt{(n+1)(n+l+\nicefrac{3}{2})}\delta_n^{n'-1} \bigg].  
\ea

We want to solve the eigenvalue problem of the operator $\hat{p}^2$ in
a finite oscillator basis truncated at an energy $(N_{\rm max}+\nicefrac{3}{2}) \, 
\hbar\omega$. For partial waves with angular momentum $l$ the basis
consists of wave functions~(\ref{waveR}) with $n=0,\ldots N$, i.e.,
the sum in Eq.~(\ref{bessel2}) is truncated at
\ba
N\equiv \left[{N_{\rm max}-l\over 2}\right] .
\ea
Here $[x]$ denotes the integer part of $x$. While $N$ clearly depends
on $l$ and $N_{\rm max}$, we will suppress this dependence in what
follows.
Motivated by Eq.~(\ref{bessel2}) we act with the matrix of $\hat{p}^2$
on the component vector $(\tilde\psi_{0,l}(k), \dots, \tilde\psi_{N,l}(k))^T$.
The well-known three-term recurrence relation for Laguerre polynomials
(see, e.g., Eq.~8.971(4) of Ref.~\cite{gradshteyn}) implies
\ba
0 &=& 
(2n+l+\nicefrac{3}{2}-b^2 k^2) \, \tilde{\psi}_{n,l}(k)
\nonumber \\
&& - \sqrt{n(n+l+\nicefrac{1}{2})} \, \tilde\psi_{n-1,l}(k) \nonumber \\
&& - \sqrt{(n+1)(n+l+\nicefrac{3}{2})} \, \tilde\psi_{n+1,l}(k) ,
\ea
and for our eigenvalue problem we arrive at
\begin{widetext}
\begin{eqnarray}
\label{eigenvalueproblem}
\hat{p}^2
\left(
\begin{array}{c}
{\tilde \psi}_{0,l}(k) \\
\vdots \\
{\tilde \psi}_{N-1,l}(k) \\
{\tilde \psi}_{N,l}(k) \end{array}
\right)
&=& k^2
\left(
\begin{array}{c}
{\tilde \psi}_{0,l}(k) \\
\vdots \\
{\tilde \psi}_{N-1,l}(k) \\
{\tilde \psi}_{N,l}(k) \end{array}
\right)
+
\left(
\begin{array}{c}
0 \\
\vdots \\
0 \\
b^{-2} \sqrt{(N+1)(N+l+\nicefrac{3}{2})} \,
 {\tilde \psi}_{N+1,l}(k) 
\end{array}
\right)
.
\label{vectorEq}
\end{eqnarray}
\end{widetext}
For $k=k_{\mu,l}$ such that 
$\tilde{\psi}_{N+1,l}(k)=0$, 
the second term on the right-hand side of Eq.~(\ref{eigenvalueproblem}) vanishes,
and we obtain an
eigenstate of the momentum operator~(\ref{p2mat}) in the finite
oscillator space.

Thus, momenta $k_{\mu,l}$ (with $\mu=0,\ldots, N$) such that
$k_{\mu,l}^2b^2$ is a root of the the Laguerre polynomial
$L_{N+1}^{l+\nicefrac{1}{2}}$ solve the eigenvalue problem of the
$\hat{p}^2$ operator in the finite oscillator space. We recall that
$L_{N+1}^{l+\nicefrac{1}{2}}$ has $N+1$ roots. Thus, in a finite model
space consisting of oscillator functions with $n=0,\ldots, N$ the
eigenvalues of the squared momentum operator are the $N+1$ roots
$k_{\mu,l}^2$ of the Laguerre polynomial
$L_{N+1}^{l+\nicefrac{1}{2}}$. We note that $k_{\mu,l}$ depends on the
angular momentum $l$ as well as $N$. To avoid a proliferation of indices, we suppress
  the latter dependence in what follows. By construction, the basis built on discrete 
momentum eigenstates is a DVR~\cite{baye1986,littlejohn2002,light2007}.

Previous studies showed that the finite oscillator basis is equivalent to a
spherical cavity at low energies~\cite{furnstahl2012}, and the radius
of this cavity is related to the wavelength of the discrete
momentum eigenstate with lowest
momentum. References~\cite{more2013,furnstahl2014} give analytical results
for the lowest momentum eigenvalue in the limit of $N\gg 1$.  The
exact determination of the eigenvalues of the momentum operator in the
present work allows us to give exact values for the radius of the
cavity corresponding to the finite oscillator basis. This radius is relevant
because it enters IR
extrapolations~\cite{coon2012,more2013,furnstahl2014}.  Let
$k_{0,l}^2b^2$ be the smallest root of the Laguerre polynomial
$L_{N+1}^{l+\nicefrac{1}{2}}$, and let $z_{0,l}$ denote the smallest
root of the spherical Bessel function $j_l$. Then 
\be
\label{Leff}
{z_{0,l}\over L(N,l)} = k_{0,l}
\ee
defines the effective radius $L$ we seek. 

The radial momentum eigenfunction corresponding to the eigenvalue $k_{\mu,l}$
in the partial wave with angular momentum $l$ has an expansion of the
form
\be
\label{discrete}
\phi_{\mu, l}(r) \equiv c_{\mu, l}\sum_{n=0}^N \tilde{\psi}_{n,l}(k_{\mu,l})\psi_{n,l}(r) .
\ee
This wave function is the projection of the spherical Bessel
  function~(\ref{bessel2}) onto the finite oscillator space. It is
  also an eigenfunction of the momentum operator projected onto the
  finite oscillator space because the specific values of $k_{\mu,l}$
  decouple this wave function from the excluded space.  In
Eq.~(\ref{discrete}) $c_{\mu, l}$ is a normalization constant that we
need to determine. In order to do so we consider the overlap
\ba
\label{cd}
\lefteqn{\langle\phi_{\mu,l}|\phi_{\nu,l}\rangle = \int\limits_0^\infty {\rm d}r r^2
\phi_{\mu, l}(r)\phi_{\nu, l}(r)} \\
&&= c_{\mu, l}c_{\nu, l}\sum_{n=0}^N \tilde{\psi}_{n,l}(k_{\mu,l})\tilde{\psi}_{n,l}(k_{\nu,l})\nonumber\\
&&= c_{\mu, l}c_{\nu, l}\sqrt{(N+1)(N+l+\nicefrac{3}{2})}\nonumber\\
&&\times \frac{\tilde{\psi}_{N,l}(k_{\mu,l}) \tilde{\psi}_{N+1,l}(k_{\nu,l}) - \tilde{\psi}_{N+1,l}(k_{\mu,l}) \tilde{\psi}_{N,l}(k_{\nu,l})}{(k_{\mu,l}^2-k_{\nu,l}^2)b^2} \nonumber .
\ea
Here, we used the Christoffel-Darboux formula for orthogonal
polynomials, see, e.g., Eq.~8.974(1) of Ref.~\cite{gradshteyn}.  As
$k_{\mu,l}$ and $k_{\nu,l}$ are roots of $\tilde{\psi}_{N+1, l}$, we confirm
orthogonality. For $k_{\mu,l}=k_{\nu,l}$ we use the rule by l'Hospital and
find (with help of Eq. 8.974(2) of Ref.~\cite{gradshteyn})
\be
\label{norm}
c_{\mu, l}^{-1} = \frac{\sqrt{(N+1)(N+l+\nicefrac{3}{2})}}{k_{\mu,l}b}\tilde{\psi}_{N, l}(k_{\mu,l}) .
\ee
It is also useful to compute the overlap
\ba
\label{over}
\langle k_{\mu,l},l|\phi_{\nu, l}\rangle &=& 
\sqrt{\pi\over 2}\int\limits_0^\infty {\rm d} r r^2 j_l(k_{\mu,l} r) \phi_{\nu, l}(r)\nonumber\\
&=& \delta_\mu^\nu c_{\mu, l}^{-1} .
\ea
This overlap vanishes for $k_{\mu,l}\ne k_{\nu,l}$,  thus, the
eigenstates of the $\hat{p}^2$ operator in finite oscillator spaces
are orthogonal to the continuous momentum eigenstates when the latter
are evaluated at the discrete momenta. This exact result is very
useful for the computation of matrix elements of a potential operator
$\hat{V}$.

For arbitrary continuous momenta $k$ we obtain from Eq.~(\ref{cd})
\ba
\label{overgen}
\tilde{\phi}_{\nu,l}(k)\equiv \langle k,l|\phi_{\nu, l}\rangle &=& 
\frac{k_{\nu,l}/b}{k_{\nu,l}^2-k^2} \tilde{\psi}_{N+1,l}(k) .
\ea
The wave function~(\ref{overgen}) is the Fourier-Bessel transform of
the discrete radial momentum wave function $\phi_{\nu, l}(r)$. 
We note that Eq.~(\ref{discrete}) relates the discrete momentum
eigenfunctions to the oscillator eigenstates via an orthogonal
transformation, implying
\ba
\label{amazing}
\sum_{\mu=0}^N c_{\mu, l}^2\tilde{\psi}_{n,l}(k_{\mu,l})\tilde{\psi}_{n',l}(k_{\mu,l}) 
= \delta_{n'}^{n} ,
\ea
and 
\be
\label{useful}
\sum_{n=0}^N \tilde{\psi}_{n,l}(k_{\mu,l})\tilde{\psi}_{n,l}(k_{\nu,l}) 
= \delta_\mu^\nu c_{\mu, l}^{-2} .
\ee
We remind the reader that the discrete set of momenta $k_{\mu,l}$ is fixed once a particular $N$ is chosen.
Equation~(\ref{amazing}) can be used to relate oscillator basis functions to the 
discrete momentum eigenfunctions~(\ref{discrete}). Thus,
\ba
\label{phi2psi}
\psi_{n,l}(r)=\sum_{\mu=0}^N c_{\mu, l} \tilde{\psi}_{n,l}(k_{\mu,l}) \phi_{\mu,l}(r) .
\ea

\subsection{Matrix elements of interactions from EFT}
\label{MatrixElementsFromEFT}
Nucleon-nucleon ($NN$) interactions from EFT are typically available
for continuous momenta in a partial-wave basis in form of the matrix
elements $\langle k',l'|\hat{V}|k,l\rangle$. Numerical integration
techniques are used to transform these matrix elements into the
oscillator basis.  However, there is a very simple approximative
relationship between the matrix elements with continuous momenta and
the matrix elements $\langle \phi_{\nu,l'}|\hat{V}|\phi_{\mu,
  l}\rangle$ in the discrete momentum basis.  This relationship is
motivated by EFT arguments and we use it in our applications of oscillator
EFT.
We consider the matrix element
\ba
\label{matele}
\lefteqn{
\langle k', l'|\hat{V}|\phi_{\mu,l}\rangle 
= \int\limits_0^\infty {\rm d} k k^2 
\langle k',l'|\hat{V}|k, l\rangle \langle k, l|\phi_{\mu, l}\rangle } \nonumber\\
&=& {1\over 2 b^3}\int\limits_0^\infty {d} x x^{l+\nicefrac{1}{2}} e^{-x}\nonumber\\
&&\times \left(\langle k', l'|\hat{V}
\frac{| x^{\nicefrac{1}{2}}b^{-1}, l\rangle \langle x^{\nicefrac{1}{2}}b^{-1}, l|} 
{x^l e^{-x}} |\phi_{\mu,l}\rangle\right) .
\ea
Here, we introduced the dimensionless integration variable $x\equiv
b^2 k^2$ and factored out a weight function \mbox{$x^{l+\nicefrac{1}{2}}
e^{-x}$} from the integrand (given in brackets) in preparation for the
next step. We evaluate the integral using ($N+1$)-point generalized Gauss-Laguerre
quadrature based on the selected weight function.
Thus, the matrix element~(\ref{matele}) becomes
\ba
\label{mateleGaussLaguerre}
\lefteqn{\langle k', l'|\hat{V}|\phi_{\mu,l}\rangle =}\nonumber\\
& & {1\over 2 b^3}\sum_{\nu=0}^N w_{\nu,l} 
\langle k', l'|\hat{V}| x_{\nu,l}^{\nicefrac{1}{2}}b^{-1}, l\rangle 
\frac{\langle x_{\nu,l}^{\nicefrac{1}{2}}b^{-1}, l| \phi_{\mu,l}\rangle}{x_{\nu,l}^l e^{-x_{\nu,l}}} \\
& & + \Delta_{N+1} . \nonumber
\ea
Here, $x_{\nu,l}$ are the roots of the Laguerre polynomial
$L_{N+1}^{l+\nicefrac{1}{2}}$, the weights are 
\ba
\label{weight}
w_{\nu,l}&\equiv& \frac{\Gamma(N+l+\nicefrac{5}{2}) x_{\nu,l}}{(N+1)!\left[(N+2)L_{N+2}^{l+\nicefrac{1}{2}}(x_{\nu,l})\right]^2}\nonumber\\
&=& \frac{\Gamma(N+l+\nicefrac{3}{2}) x_{\nu,l}}{(N+1)!(N+l+\nicefrac{3}{2})\left[L_{N}^{l+\nicefrac{1}{2}}(x_{\nu,l})\right]^2} , 
\ea
and the error term is
\be
\label{error}
\Delta_{N+1} \equiv \frac{(N+1)! \Gamma(N+l+\nicefrac{5}{2})}{(2N+1)!}f^{(2N+2)}(\xi) ,
\ee
see, e.g., Ref.~\cite{concus1963}. 
For the weights, we also used Eq.~8.971(6) of
Ref.~\cite{gradshteyn}. 
In the error term, $f^{(2N+2)}(\xi)$ denotes the $(2N+2)$-th derivative of the integrand (given in round brackets) of Eq.~(\ref{matele}), evaluated at  $\xi$ which is somewhere in the
integration domain. 

We want to estimate the order of the error term $\Delta_{N+1}$
when using EFT interactions.
For this purpose, we 
write the potential as a sum of separable potentials
\be
V(k',k) = \sum_a v_a g_a(k')g_a(k) , 
\ee
and write
\be
g_a(k) = (k/\Lambda)^l e^{-{1\over 2}k^2b^2} \tilde{g}_a(k/\Lambda) .
\ee
Here, $\Lambda$ is a high-momentum cutoff.  The function
$\tilde{g}_a(k/\Lambda)$ is an even function of its arguments 
(see, e.g., Ref.~\cite{machleidt2011}), 
and can be expanded in a Taylor series
\be 
\label{taylor}
\tilde{g}_a(k) = \sum_{n=0}^\infty {\tilde{g}_a^{(n)}(0) \over
  n!}\left(x\over\Lambda^2 b^2\right)^n . 
\ee
Here, we again used $x\equiv b^2k^2$. With this expansion in mind, and
noting that the wave function $\tilde\phi_{\nu,l}$ 
can be expanded in terms
of oscillator wave functions, the integrand $f$
(given in round brackets) of Eq.~(\ref{matele}) is a product of
$\tilde{g}_a$ and a sum of Laguerre polynomials (from the wave
function $\tilde\phi_{\nu, l}$). The ($N$+1)-point Gauss-Laguerre integration is exact for
monomials up to $x^{2N+1}$.
As the wave function $\tilde\phi_{\nu, l}$
contains monomials up to $x^N$, the Gauss-Laguerre integration becomes
inexact for terms starting at $n=N+2$ in the Taylor
series~(\ref{taylor}). Thus, $f^{(2N+2)}$ in the error
term~(\ref{error}) scales as $1/(\Lambda b)^{2N+l+4}$.  
In the
oscillator EFT, typical momenta $k$ scale as $1/b$, and the error term scales as
\be
\Delta_{N+1} = \mathcal{O}\left(\left(k/\Lambda\right)^{2N+l+4}\right) .
\ee
Therefore,  
\ba
\langle k', l'|\hat{V}|\phi_{\mu,l}\rangle &=& 
\langle k',l'|\hat{V}|k_{\mu,l},l\rangle c_{\mu, l} \nonumber\\
&&+\mathcal{O}\left(\left(k/\Lambda\right)^{2N+l+4}\right) .
\ea
Repeating the calculation for the bra side yields the final result
\ba
\label{final}
\langle \phi_{\nu, l'}|\hat{V}|\phi_{\mu, l}\rangle 
&=& c_{\nu, l'} c_{\mu, l}\langle k_{\nu,l'}, l' |\hat{V}|k_{\mu,l},l\rangle 
\nonumber\\
&&+\mathcal{O}\left(\left(k/\Lambda\right)^{2N+l+4}\right) .
\ea
In oscillator EFT, we will omit the correction term and set 
\ba
\label{Vmain}
\langle \phi_{\nu, l'}|\hat{V}|\phi_{\mu, l}\rangle 
&=& c_{\nu, l'} c_{\mu, l}\langle k_{\nu,l'}, l' |\hat{V}|k_{\mu,l},l\rangle .
\ea
We note that this assignment seems to be very natural for an EFT
built on a finite number of discrete momentum states.
For sufficiently large $N$, the difference to the matrix element
obtained from an exact integration can be view as a correction that is
beyond the order of the power counting of the EFT we build upon. In
Eq.~(\ref{Vmain}) the matrix elements between the discrete and
continuous momentum states simply differ by normalization factors
because of the different normalization~(\ref{over}) of discrete and
continuous momentum eigenstates. Thus, partial-wave decomposed matrix
elements of any momentum-space operator can readily be used to compute
the corresponding oscillator matrix elements. In the
Appendix~\ref{appendix} we present an alternative motivation for the
usage of Eq.~(\ref{Vmain}).

In the remainder of this Subsection, we give useful formulas that
relate the matrix elements and wave functions of the discrete momentum
basis and the oscillator basis.  We note that the oscillator basis states
$|\psi_{n,l}\rangle$ are related to the discrete momentum states
$|\phi_{n,l}\rangle$ via Eq.~(\ref{phi2psi}).  Thus, we can also give
an useful formula that transforms momentum-space matrix elements to
the oscillator basis according to
\ba
\label{HO_V}
\lefteqn{\langle \psi_{n', l'}|\hat{V}|\psi_{n, l}\rangle 
=}\\ &&\sum_{\nu,\mu=0}^N c_{\nu, l'}^2 \tilde{\psi}_{n',l'}(k_{\nu,l'})
\langle k_{\nu,l'}, l' |\hat{V}|k_{\mu,l},l\rangle c_{\mu, l}^2 \tilde{\psi}_{n,l}(k_{\mu,l}) 
\nonumber\\
&&+ \mathcal{O}\left(k^{2N+2}\right) .
\ea 
This formula also reflects the well known fact that the oscillator basis mixes
low- and high-momentum physics. 

The relation between matrix elements in the oscillator basis and the discrete
momentum basis is given by
\ba
\lefteqn{\langle\phi_{\nu,l'}|\hat{V}|\phi_{\mu,l}\rangle =}\nonumber\\
&&c_{\nu, l'}c_{\mu, l}
\sum_{n, n'=0}^N \tilde{\psi}_{n',l'}(k_{\nu,l'}) \langle \psi_{n',l'}|\hat{V}|\psi_{n,l}\rangle \tilde{\psi}_{n,l}(k_{\mu,l}) .
\ea

Finally, we discuss the inversion of Eq.~(\ref{HO_V}), e.g.,
for situations where scattering processes in the continuum have to
be considered for interactions based on oscillator spaces. We obtain
\ba
\label{2mom}
\lefteqn{\langle k_{\nu,l'}, l' |\hat{V}|k_{\mu,l},l\rangle =}\nonumber\\ 
&&\sum_{n, n'=0}^N\tilde{\psi}_{n',l'}(k_{\nu,l'}) 
\langle \psi_{n', l'}|\hat{V}|\psi_{n, l}\rangle \tilde{\psi}_{n,l}(k_{\mu,l}) \\
&& +\mathcal{O}\left(k^{2N+l+4}\right) , \nonumber
\ea
because of Eq.~(\ref{useful}).

For arbitrary momenta, one needs to use the overlaps~(\ref{overgen})
in the evaluation of the matrix elements, and finds the generalization
of Eq.~(\ref{2mom}) as
\ba
\label{Vproj}
\lefteqn{\langle k', l' |\hat{\Pi}\hat{V}\hat{\Pi}|k,l\rangle =
\tilde{\psi}_{N+1,l'}(k')\tilde{\psi}_{N+1,l}(k) }\nonumber\\ 
&&\times\sum_{\nu, \mu = 0}^N
\frac{k_{\nu,l'}/b}{(k')^2-k_{\nu,l'}^2} \langle \phi_{\nu, l'}|\hat{V}|\phi_{\mu, l}\rangle
\frac{k_{\mu,l}/b}{k^2-k_{\mu,l}^2} .
\ea
Here, we introduced the projection operator onto the finite oscillator space
\be
\label{projector}
\hat{\Pi}\equiv \sum_{\nu = 0}^N \sum_{l=0}^{N_{\rm max}-2\nu} |\phi_{\nu,l}\rangle\langle\phi_{\nu,l}| .
\ee 
We note that the projection operator acts as a UV (and IR)
regulator. It is nonlocal, and can be written in many ways. Examples
are
\ba
\lefteqn{
\langle k', l|\hat{\Pi}|k,l\rangle 
= \sum_{n=0}^N \tilde{\psi}_{n,l}(k')\tilde{\psi}_{n,l}(k) } \\
&=& {\sqrt{(N+1)(N+l+\nicefrac{3}{2})}\over b^2}\nonumber\\
&\times& \frac{\tilde{\psi}_{N,l}(k) \tilde{\psi}_{N+1,l}(k') - \tilde{\psi}_{N+1,l}(k) \tilde{\psi}_{N,l}(k')}{k^2-(k')^2} \\
&=&{\tilde{\psi}_{N+1,l}(k')\tilde{\psi}_{N+1,l}(k) \over b^2} \nonumber\\
&&\times\sum_{\nu=0}^N
\frac{k_{\nu,l}^2}{[(k')^2-k_{\nu,l}^2][k^2-k_{\nu,l}^2]} .
\label{proj38}
\ea
Here, the first identity comes directly  from the definition of the
projector in terms of the oscillator eigenfunctions. The second
identity follows from the calculation displayed in Eq.~(\ref{cd}),
while the third identity follows from Eq.~(\ref{overgen}).

This presents us with an alternative motivation (but not derivation)
of Eq.~(\ref{Vmain}). We evaluate the projected matrix elements~(\ref{Vproj})
at discrete momenta and find with Eq.~(\ref{overgen}) that
\ba
\langle k_{\nu, l'},l'|\hat{\Pi}\hat{V}\hat{\Pi}|k_{\mu, l}, l\rangle 
= c_{\nu, l'}^{-1} c_{\mu, l}^{-1}\langle \phi_{\nu,l'} |\hat{V}|\phi_{\mu,l}\rangle .
\ea  
We note that 
\be
\langle k_{\nu, l'},l'|\hat{\Pi}\hat{V}\hat{\Pi}|k_{\mu, l}, l\rangle 
=\langle k_{\nu, l'},l'|\hat{V}|k_{\mu, l}, l\rangle 
\ee
approximately holds for the discrete momenta in the finite oscillator
space (cf. Eq.~(\ref{final})).


\section{Chiral interactions in finite oscillator bases}
\label{optimize}

In this Section we present a proof-of-principle construction of a chiral
$NN$ interaction in the framework of
the oscillator EFT. First, we study the effects that the truncation to
a finite oscillator basis has on phase shifts of existing $NN$
interactions. 
Second, we demonstrate that a momentum-space chiral interaction at NLO
can be equivalently constructed in oscillator EFT.

We consider the
chiral interactions N$^3$LO$_{\rm EM}$~\cite{entem2003} and
NLO$_{\rm sim}$~\cite{carlsson2015}.
Both
interactions employ regulators of the form
\ba
\label{regulator}
f(q)=\exp\left[-\left({q\over\Lambda_\chi}\right)^{2n}\right] . 
\ea
Here, $q$ is a relative momentum, $n$ is an integer, and $\Lambda_\chi$ is
the high-momentum cutoff,
specifically $\Lambda_\chi =500$~MeV. We use $n=3$ in what follows.
This cutoff needs to be
distinguished from the (hard) UV cutoff~\cite{koenig2014}
\be
\label{uv}
\Lambda_{\rm UV} \approx \sqrt{2\left(N_{\rm max}
  +\nicefrac{7}{2}\right)}/b \ee
of the oscillator-EFT interaction. 

 Let us comment on using the projector~(\ref{projector}) in
  combination with the regulator~(\ref{regulator}).  In momentum
  space, the projector~(\ref{projector}) is approximately the identity operator for
  momenta $k,k'$ between the IR and UV cutoffs $\Lambda_{\rm IR}$ and
  $\Lambda_{\rm UV}$ of the oscillator basis. For
  momenta $k,k'> \Lambda_{\rm UV}$ the projector~(\ref{projector})
  falls off as a Gaussian. For the regulator~(\ref{regulator})
  we choose $\Lambda_\chi < \Lambda_{\rm UV}$, which introduces a
  super-Gaussian falloff
  for momenta
  $\Lambda_\chi\lesssim q \lesssim\Lambda_{\rm UV}$.
  As an example, let us consider $\Lambda_\chi
  =500$~MeV and $\Lambda_{\rm UV}=700$~MeV. Then, $f(\Lambda_{\rm
    UV})\approx 5\times 10^{-4}$ at the point where the super-Gaussian
  falloff goes over into a Gaussian falloff. Assuming a ratio
  $q/\Lambda_{\chi}=\nicefrac{1}{3}$ that is typical for the power counting in chiral EFT,
  $f(\Lambda_{\rm UV}) \approx (\nicefrac{1}{3})^7$, and the asymptotic Gaussian
  falloff is not expected to introduce significant contributions to
  contact interactions at NLO. Eventually, one might want to consider
  removing the regulator from an oscillator-based EFT. At this moment
  however, it is also useful in taming the oscillations discussed
  below and shown in Fig.~\ref{EntemMachleidt_oscillations}.

\subsection{Effects of finite oscillator spaces on phase shifts}

\begin{figure}[tb]
\includegraphics[scale=0.62]{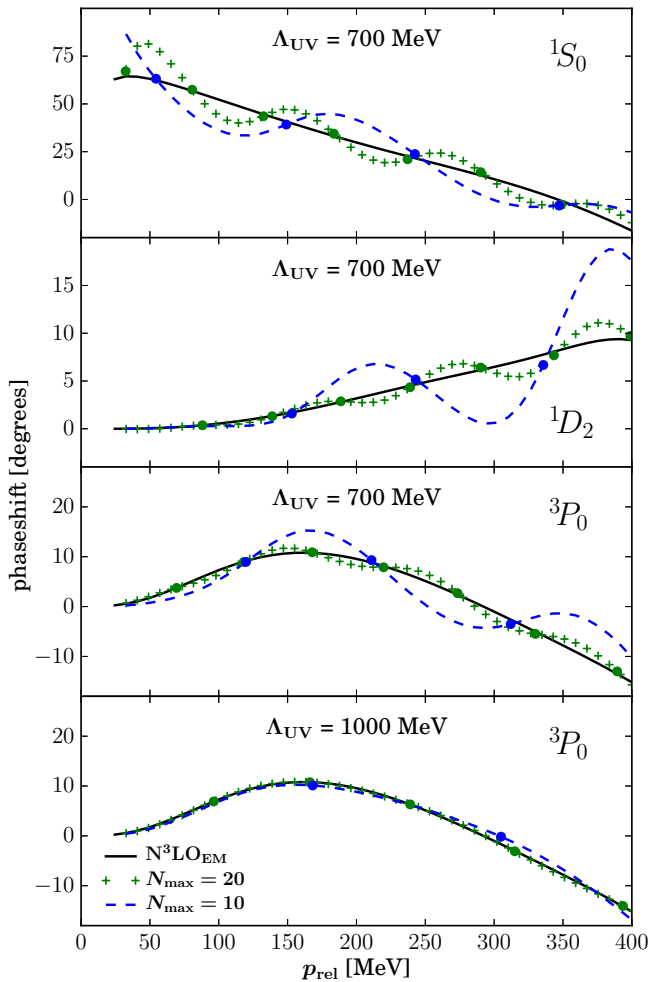}
\caption{ (Color online) Computed $np$ phase shifts of the $NN$
  potential N$^3$LO$_{\rm EM}$~\cite{entem2003} projected onto finite
  oscillator spaces with $N_{\rm max}$ = 10 and $N_{\rm max}$ = 20 compared to the phase
  shifts from the full (i.e., not projected) interaction in selected
  partial waves as a function of the relative momentum. In the top
  three panels the oscillator frequencies were chosen as \mbox{$\hbar
  \omega$ = 40~MeV} and 23~MeV, respectively, in order to obtain an
  oscillator cutoff $\Lambda_{\rm UV} \approx 700$~MeV that well
  exceeds the chiral cutoff $\Lambda_{\chi} = 500$~MeV.  The filled
  circles mark the phase shifts at the momenta corresponding to the
  eigenenergies of the scattering channel in the truncated oscillator
  space.
In the bottom panel $\hbar\omega$ is increased to 80~MeV (for \mbox{$N_{\rm
  max}=10$}) and 46~MeV (for $N_{\rm max}=20$), respectively. This yields
a UV cutoff $\Lambda_{\rm UV} \approx 1000$ MeV in the oscillator
basis and significantly reduces the phase shift oscillations.
 }
\label{EntemMachleidt_oscillations}
\end{figure}

It is instructive to study the effects that a projection onto a finite
oscillator basis has on phase shifts.  For this purpose, we employ the
well known $NN$ interaction N$^3$LO$_{\rm
  EM}$~\cite{entem2003}. First, we transform its matrix elements to a
finite oscillator space using numerically exact quadrature, and
subsequently compute the phase shifts using the $J$-matrix
  approach of Ref.~\cite{shirokov2004}.  The two parameter
combinations $N_{\rm max} = 10$, $\hbar\omega = 40$~MeV, and $N_{\rm
  max}= 20$, $\hbar\omega = 23$~MeV respectively yield a UV cutoff
$\Lambda_{\rm UV}\approx 700$~MeV, see Eq.~(\ref{uv}). This
considerably exceeds the high-momentum cutoff $\Lambda_\chi$ of the
interaction.
Figure~\ref{EntemMachleidt_oscillations} shows
the resulting phase shifts in selected partial waves.  
%
%
%
%
The projection onto finite oscillator bases introduces oscillations in
the phase shifts.
On the one hand, many {\it ab initio} calculations of atomic nuclei
yield practically converged results for bound-state observables in
model spaces with $N_{\rm max} \lesssim 14$ and oscillator frequencies
around $24$~MeV.
On the other hand, oscillator spaces consisting of only 10 to 20
shells are clearly too small to capture all the information contained
in the original interaction.
We note (i) that the N$^3$LO$_{\rm EM}$ interaction is formulated for
arbitrary continuous momenta whereas oscillator EFT limits the
evaluation to a few discrete momenta, and (ii) that the present
$\Lambda_{\rm UV}$ cuts off the high-momentum tails of the
N$^3$LO$_{\rm EM}$ interaction.

For $N_{\rm max}\rightarrow\infty$ we expect to arrive at the original
phase shifts, and this is supported in
Fig.~\ref{EntemMachleidt_oscillations} by the reduced oscillations in
the $N_{\rm max}= 20$ phase shifts compared to their $N_{\rm max} =
10$ counterparts. 
The period of oscillations is approximately given by
the IR cutoff.
In the bottom panel of Fig.~\ref{EntemMachleidt_oscillations}, the
oscillator spacing $\hbar\omega$ is increased to 80~MeV (for $N_{\rm
  max}=10$) and 46~MeV (for $N_{\rm max}=20$), respectively, yielding
a UV cutoff $\Lambda_{\rm UV} \approx 1000$ MeV in the oscillator
basis.  The phase shift oscillations are significantly reduced for
such large values of $\Lambda_{\rm UV}$.

We note that the eigenvalues of the Hamiltonian matrix in \mbox{$(N_{\rm
  max}+1)$} oscillator shells play a special role for the phase shifts
in uncoupled partial wave channels. At these energies, the
eigenfunctions are standing waves with a Dirichlet boundary condition
at the (energy-dependent) radius of the spherical cavity that is
equivalent to the finite oscillator basis, and one can alternatively
use this information in the computation of the phase
shifts~\cite{luu2010,more2013}.
The filled circles in Fig.~\ref{EntemMachleidt_oscillations} indicate
the values of the phase shifts at these energies, which are close to
those of the original interaction.

{ 
Subsection~\ref{MatrixElementsFromEFT} discusses two ways of projecting
 momentum-space interactions onto a finite oscillator
space.
The first approach involves the determination of the
matrix element~(\ref{matele}) via an exact numerical integration over
continuous momenta.
The second approach, Eq.~(\ref{Vmain}),
uses $(N+1)$-point Gauss-Laguerre quadrature
to compute the integral in Eq.~(\ref{matele}). This only
requires us to evaluate the interaction at those momenta
that are physically realized in the finite oscillator
basis, which is more in the spirit of an EFT.
Because we want to follow the oscillator EFT approach in later
sections, we study the effect that the error term~(\ref{error})
associated with the $(N+1)$-point Gauss-Laguerre quadrature has on the
projected phase shifts.
Figure~\ref{EntemMachleidt_projectionType} shows a comparison of projected 
N$^3$LO$_{\rm EM}$ phase shifts obtained
from the two projection approaches to the
original N$^3$LO$_{\rm EM}$ phase shifts.
Overall, both versions yield very similar phase shifts.
For the phase shifts associated with the $(N+1)$-point Gauss-Laguerre
quadrature, the oscillations seem to be somewhat
reduced for small energies.
Also, we find a notably improved agreement between the $(N+1)$-point
Gauss-Laguerre phase shifts and the original ones at the discrete
eigenenergies of the scattering channel Hamiltonians, as indicated by
the full circles.
From now on we exclusively use the $(N+1)$-point Gauss-Laguerre
integration to compute matrix elements in oscillator EFT.
%

%
%
\begin{figure}[tb]
\includegraphics[scale=0.62]{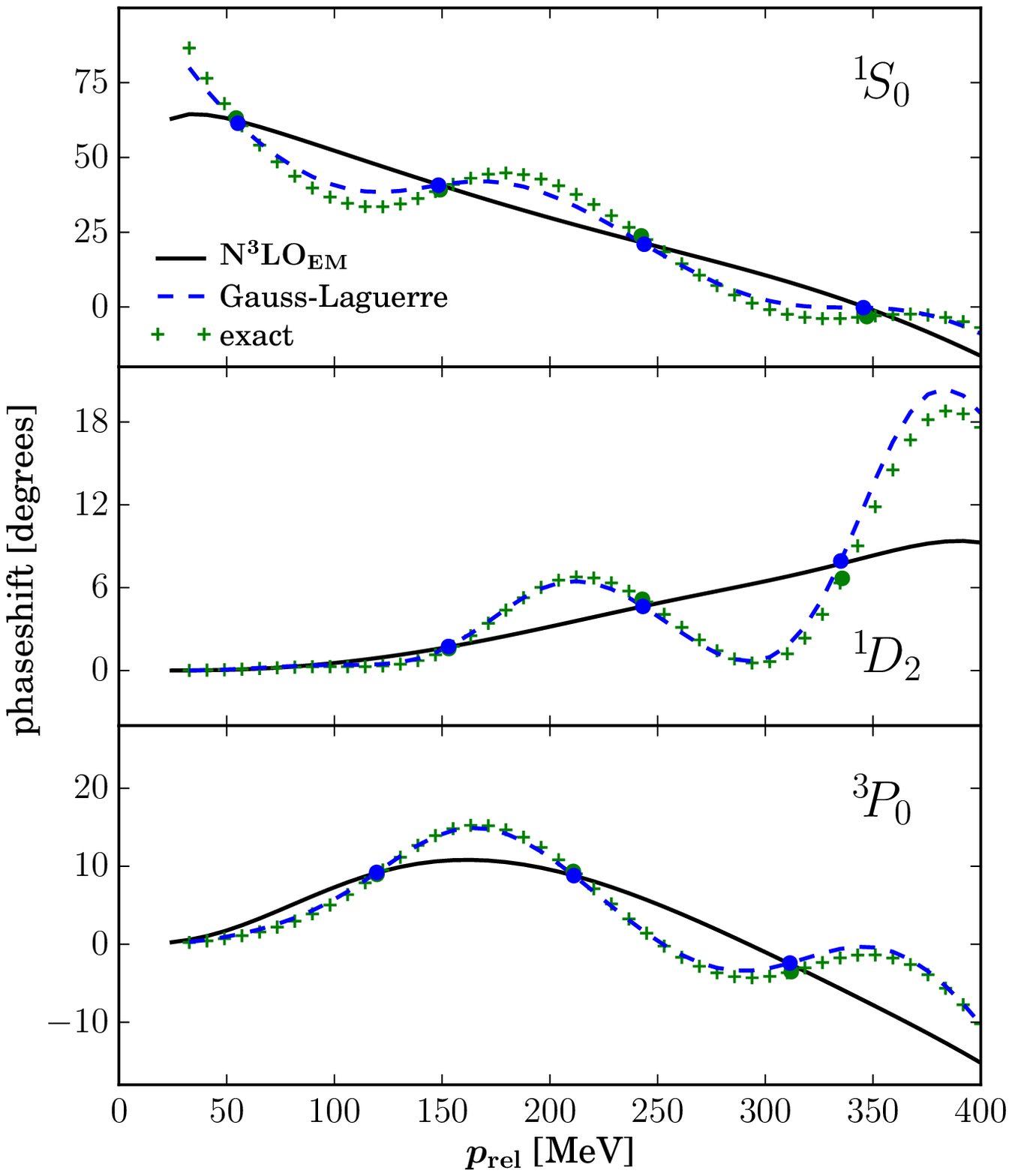}
\caption{ (Color online) 
Phase shifts of N$^3$LO$_{\rm EM}$ 
obtained from a projection onto a 
\mbox{$N_{\rm max} = 10$}, 
\mbox{$\hbar\omega = 40$ MeV}
oscillator space using 
(i) exact integration in Eq.~(\ref{matele}) (blue, dashed line)
and 
(ii) Gauss-Laguerre quadrature Eq.~(\ref{Vmain}) (green crosses),
compared to the unprojected N$^3$LO$_{\rm EM}$ phase shifts (black, solid line).
The filled
  circles mark the phase shifts at the momenta corresponding to the
  eigenenergies of the scattering channel in the truncated oscillator
  space.
}
\label{EntemMachleidt_projectionType}
\end{figure}
%
%

\subsection{Reproduction of the phase shifts of a NLO interaction}

In what follows we employ the oscillator EFT at NLO. All matrix
elements in oscillator EFT are based on Eq.~(\ref{Vmain}), i.e., the
matrix elements from continuum momentum space are evaluated at the
discrete momenta of the finite oscillator basis. At this order the
chiral interactions depend on 11 LECs and exhibit sufficient
complexity to qualitatively describe nuclear properties.
Specifically, in this Section we aim at reproducing the phase shifts
and selected deuteron properties of the chiral interaction NLO$_{\rm
  sim}$~\cite{carlsson2015} by optimizing these 11 LECs.
Throughout this work, the $\chi^2$ fits evaluate the phase shifts at
20 equidistant energies in the laboratory energy range up to 350 MeV,
with weights $\sigma^2 \propto (q/ \Lambda_\chi)^3$. We note that
  more sophisticated weights (including also the pion mass or the
  oscillatory patterns) would be needed for a quantification of
  uncertainties, see,
  e.g. Refs.~\cite{stump2001,carlsson2015,furnstahl2015a}. In this
  work we only investigate the feasibility of an oscillator EFT with
  regard to the computation of heavy nuclei.

The main goal of oscillator EFT is to enable the computation
of heavy nuclei. Therefore, we set $N_{\rm max} = 10$ for the interaction. This allows us to perform IR extrapolations based on calculations in spaces with $N_{\rm max} = 10$, 12, and 14  for the kinetic energy.
To determine $\hbar\omega$  we study phase shifts in selected partial waves in Fig.~\ref{NLO_oscillations}.
The first value, $\hbar\omega=20$~MeV, again
corresponds to a UV cutoff of $\Lambda_{\rm UV}\approx 500$~MeV and
yields the familiar oscillations in the phase shifts. The second value,
$\hbar\omega=34$~MeV, corresponds to a UV cutoff $\Lambda_{\rm UV}\approx
650$~MeV that significantly exceeds the chiral cutoff
$\Lambda_{\chi}\approx 500$~MeV. In this case, the oscillations are
drastically reduced. Consequently, we use $\hbar\omega = 34$ MeV in the following.

\begin{figure}[tb]
\includegraphics[scale=0.62]{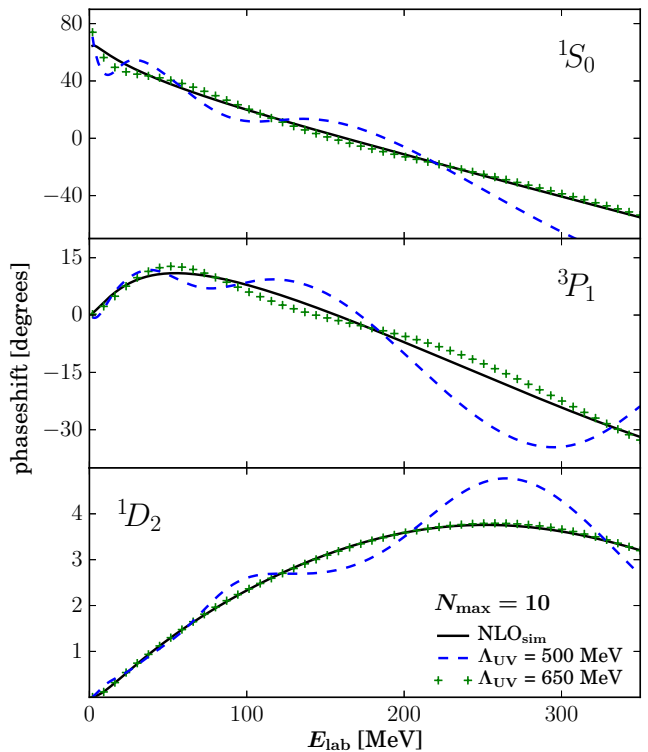}
\caption{(Color online) Phase shifts of NLO interactions obtained from
  oscillator EFT through fits to $\text{NLO}_\text{sim}$ phase shifts
  compared to the $\text{NLO}_\text{sim}$ data~\cite{carlsson2015}.
The momentum-space interaction matrix elements used in the
construction of the oscillator EFT interactions employ a fixed chiral
cutoff $\Lambda_\chi = 500$ MeV.
We fix $N_{\rm max} = 10$ and consider the two oscillator frequencies
 $\hbar\omega = 20$~MeV and 34~MeV, yielding UV cutoffs 
 $\Lambda_{\rm UV} = 500$~MeV and 650~MeV, respectively.
For $\Lambda_{\rm UV} = 650$~MeV,  the oscillations in the 
phase shifts are notably reduced.  }
\label{NLO_oscillations}
\end{figure}
%
%
 
In Fig.~\ref{NLOsim_phaseshifts} we compare further phase shifts from the oscillator EFT to the $\text{NLO}_\text{sim}$ phase shifts.  We note that the channel
${^1\!D_2}$ is a prediction because there is no corresponding LEC at this chiral order.  
%
\begin{figure}[p]
\includegraphics[scale=0.58]{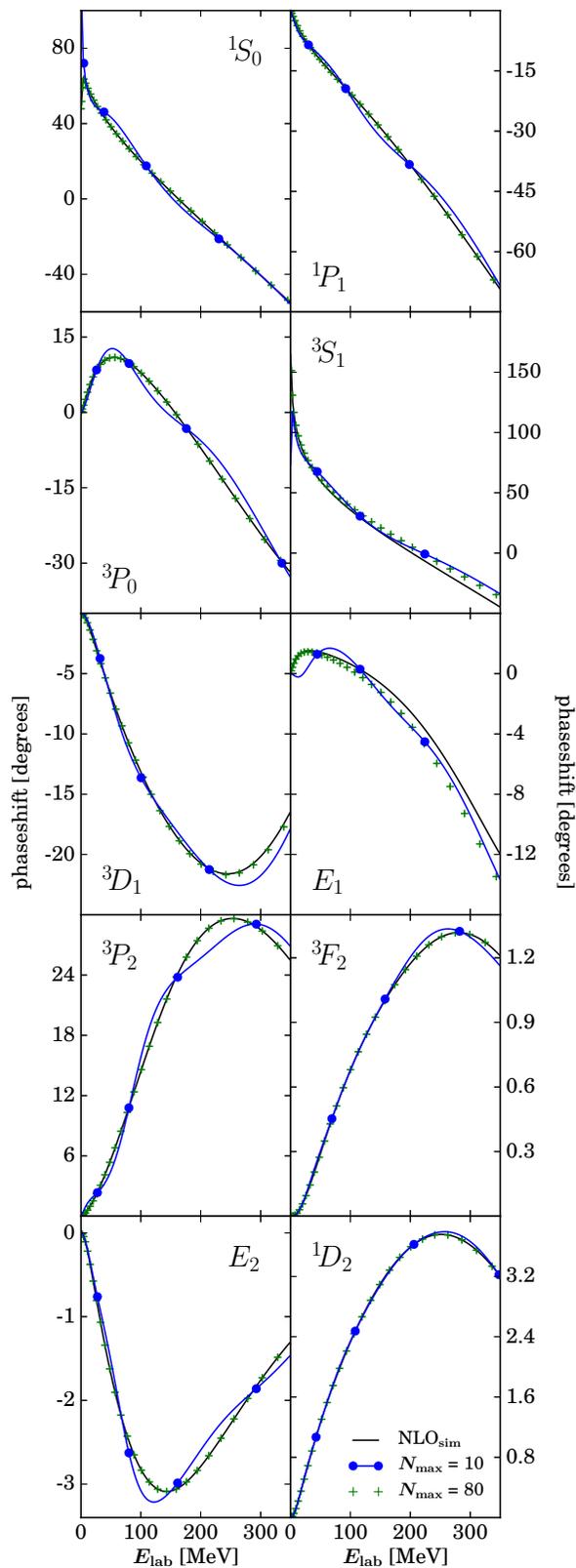}
\caption{(Color online) Phase shifts from oscillator EFT with \mbox{$N_{\rm
    max}=10$} (full blue circles and blue line) compared to NLO$_{\rm
    sim}$ phase shifts (black line) for partial wave channels as
  indicated. Also shown (green crosses) are results from oscillator
  EFT in a large model space with \mbox{$N_{\rm max}=80$}.
  }
\label{NLOsim_phaseshifts}
\end{figure}
%
For completeness, and to assess the large-$N_{\rm max}$ behavior, we also show the results of an optimization in oscillator EFT with $N_{\rm max} = 80$ and $\hbar\omega = 6$ MeV ($\Lambda_{\rm UV} \approx 650$~MeV) as green crosses.
In general, oscillator EFT reproduces the phase shifts  well
over the whole energy range up to the pion-production threshold.

\begin{table}[bt]
\begin{tabular}{| l d d d | }
\hline \hline
        & \multicolumn{1}{c}{\hspace{10pt} $N_{\rm max}=10$} & \multicolumn{1}{c}{\hspace{10pt} $N_{\rm max}=80$}   &  \multicolumn{1}{c|}{\hspace{10pt} $\rm{NLO}_{\rm sim}$}   \\
\hline
$E_d \ [\text{MeV}]$ & -2.268 & -2.212 & -2.225 \\
$r_d \ \ [\text{fm}]$     & 2.108 & 1.962 & 1.976 \\
$Q_d\ [\text{fm}^2]$ & 0.195 & 0.259 & 0.259 \\
$P_d$                      & 0.028 & 0.029 & 0.029 \\
\hline
$\tilde{C}^{(np)}_{^1\!S_0}$ & -0.157499 & -0.150552 & -0.150623 \\
$\tilde{C}^{(pp)}_{^1\!S_0}$ & -0.15513   & -0.14877   & -0.14891 \\
$\tilde{C}^{(nn)}_{^1\!S_0}$ & -0.15645  & -0.14990    & -0.14991 \\
$C_{^1\!S_0}$                      &  1.7972    & 1.6910       & 1.6935 \\
$\tilde{C}_{^3\!S_1}$            &  -0.1721   & -0.1752     & -0.1843 \\
$C_{^3\!S_1}$                      & -0.406      & -0.389      & -0.218 \\ 
$C_{E_1}$                         &  0.224       & 0.232        & 0.263 \\
$C_{^3\!P_0}$                      &  1.3038    & 1.2995       & 1.2998 \\
$C_{^1\!P_1}$                      &  1.047      & 1.023         & 1.025 \\
$C_{^3\!P_1}$                      &  -0.335     & -0.336       & -0.336 \\
$C_{^3\!P_2}$                      &  -0.2010   & -0.2028     & -0.2029 \\
 \hline \hline
\end{tabular}
\label{tba}
\caption{
Comparison of deuteron properties and LECs  
for effective interactions with \mbox{$N_{\rm max} = 10$}, 80 obtained from a fit to NLO$_{\rm sim}$ phase shifts and deuteron properties, compared to NLO$_{\rm sim}$ data~\cite{carlsson2015}.
The deuteron properties, computed in large $N_{\rm max}=100$ model spaces, are the ground-state energy $E_d$, point-proton radius $r_d$, quadrupole moment $Q_d$, and $D$-state probability $P_d$.
The oscillator frequencies $\hbar\omega$ are 34 MeV and
6 MeV, respectively, corresponding to a UV cutoff of \mbox{$\Lambda_{\rm UV}\approx$ 650 MeV}.
Units of the LECs as in Ref.~\cite{machleidt2011}.
}
\label{NLOsimTable}
\end{table}
%
%

In the deuteron channel we fit not only to phase shifts but also to
its binding energy $E_d$, point-proton radius $r_d$, and quadrupole
moment $Q_d$.  For \mbox{$N_{\rm max}=80$}, we reproduce the deuteron
properties well, as can be seen in Table~\ref{NLOsimTable}.  
For the \mbox{$N_{\rm max}=10$} effective interaction, it becomes more difficult
to simultaneously reproduce all data. 
Therefore, we relax the
requirement to reproduce the quadrupole moment in favor of the other
deuteron properties and the phase shifts.
  We converge the deuteron calculations by employing large
model spaces
of 100 oscillator shells
 where only the kinetic energy acts beyond the \mbox{$N_{\rm max}$} 
used to define the effective interaction.

In Table~\ref{NLOsimTable} we also compare the LECs of the effective
interactions for \mbox{$N_{\rm max}$ = 10} and 80 to the LECs of NLO$_{\rm
  sim}$.
For $N_{\rm max}$ = 80, our approach quite
accurately reproduces the NLO$_{\rm sim}$ LECs, with $C_{^3\!S_1}$ being
the only exception. 
Because this LEC is associated with the deuteron
channel, its value is affected by the weighting of the deuteron
properties in the fit, and  we expect that a better reproduction can be achieved by assigning different weights to the deuteron 
properties.
More importantly, with the exception of $C_{^3\!S_1}$, the values of
the LECs for the \mbox{$N_{\rm max}$ = 10} interaction are very
similar to the LECs of the NLO$_{\rm sim}$ interaction. Also, all of
the values are of natural size.

Let us also address the sensitivity of our results to the UV cutoff of
the employed oscillator space. For this purpose we keep
$\hbar\omega=34$~MeV fixed and optimize two more interactions defined
in model spaces with $N_{\rm max}=12$ and $N_{\rm max}=14$,
respectively. We recall that the UV cutoff of the model space
increases with increasing $N_{\rm max}$, and that $N_{\rm
  max}=10,12,14$ corresponds to UV cutoff \mbox{$\Lambda_{\rm UV}
  \approx$ 650, 700, 750 MeV}, respectively.  Resulting few-body
observables from these interactions are shown in
Table~\ref{UVconvergence}, and also compared to the infinite-space
interaction $\rm NLO_{sim}$. We see that results converge slowly
toward those of the $\rm NLO_{sim}$ interaction as $N_{\rm max}$ is
increased. We also note that the few-body observables from the
different $N_{\rm max}$ interactions exhibit differences that are
consistent with uncertainty expectations at
NLO~\cite{epelbaum2015,lynn2016,carlsson2015}.
 
\begin{table}[tb]
\begin{tabular}{|lcccc|}
\hline \hline 
 & $N_{\rm max} = 10$ \ & \ $N_{\rm max} = 12$ \ & \ $N_{\rm max} = 14$ \ & \ $\rm NLO_{sim}$ \\ 
\hline 
$E_d$ & -2.261 & -2.227 & -2.225 & -2.224 \\ 
$r_d$ & 2.108 & 1.974 & 1.889 & 1.975 \\ 
$Q_d$ & 0.195 & 0.279 & 0.272 & 0.259 \\ 
$P_d$ & 0.028 & 0.031 & 0.031 & 0.028 \\ 
$E_{^3\text{H}}$ & -8.944 & -8.502 & -8.094 & -8.270 \\ 
$r_{^3\text{H}}$ & 1.627 & 1.592 & 1.578 & 1.614 \\ 
$E_{^3\text{He}}$ & -8.169 & -7.735 & -7.33 & -7.528 \\ 
$r_{^3\text{He}}$ & 1.787 & 1.764 & 1.761 & 1.791 \\ 
$E_{^4\text{He}}$ & -28.736 & -27.96 & -27.302 & -27.44 \\ 
$r_{^4\text{He}}$ & 1.465 & 1.463 & 1.457 & 1.482 \\ 
\hline \hline 
\end{tabular}
\caption{ Deuteron properties as in Table~\ref{NLOsimTable}, as well
  as binding energies (in MeV) and radii (in fm) for ${}^3\rm H$, ${}^3\rm
  He$, and ${}^4\rm He$ from different effective interactions,
  compared to $\rm NLO_{sim}$.  The oscillator length $\hbar\omega$ is
  fixed at 34~MeV, and we consider \mbox{$N_{\rm max}$ = 10, 12, 14},
  corresponding to values of the UV cutoff \mbox{$\Lambda_{\rm UV}
    \approx$ 650, 700, 750 MeV}.  }
\label{UVconvergence}
\end{table}

Clearly, the NLO interaction from oscillator EFT differs from
NLO$_{\rm sim}$ through the complicated projection that introduces IR
and UV cutoffs and is highly nonlocal, see Eq.~(\ref{projector}). 
In the EFT sense
the difference between these interactions should
be beyond the order at which we are currently operating.
 While we cannot prove this
equivalence, the numerical results of this Section encourage us to
pursue the construction of a chiral NLO interaction within oscillator
EFT by optimization to the phase shifts from a high-precision $NN$
potential in the following Section.


\section{NLO interactions in oscillator EFT and many-body results}
\label{results}

In what follows, we set \mbox{$N_{\rm max} = 10$} for the interaction. On the one hand,
lower oscillator frequencies correspond to larger oscillator lengths
and lead to a rapid IR convergence.  On the other hand, lower
oscillator frequencies also correspond to lower UV cutoffs.  In what
follows we choose $\hbar\omega = 24$~MeV which results in a UV cutoff
$\Lambda_{\rm UV}\approx 550$~MeV and an IR length $L\approx 9.6$~fm
according to Eq.~(\ref{Leff}).
Considering the tail of the regulator function~(\ref{regulator}) we
set $\Lambda_{\chi}=450$~MeV, which ensures that $\Lambda_{\rm UV}$
significantly exceeds $\Lambda_{\chi}$.

Practical calculations are performed in the laboratory system, 
and use
the $N_{\rm max}=10$ interaction together with the intrinsic kinetic
energy in  oscillator spaces with $N^{\rm lab} = 2n_{1} + l_1 \ge N_{\rm max}$. Here $n_1$ and $l_1$ are the radial and angular momentum
quantum numbers of the harmonic oscillator in the laboratory
frame. 
Results for $N^{\rm lab}=10, 12, 14$ are feasible  and will allow us to perform IR
extrapolations for bound-state energies and radii. 

In this Section, we construct a chiral interaction at NLO from
realistic phase shifts, and subsequently utilize it in coupled-cluster
calculations of $^4$He, $^{16}$O, $^{40}$Ca, $^{90}$Zr, and
$^{132}$Sn. Our main objectives are (i) to present a
proof-of-principle optimization of a realistic interaction within the
framework of the oscillator EFT, and (ii) to demonstrate that such an
interaction converges fast even in heavy nuclei.  We
compute the matrix elements in oscillator EFT, i.e., based on
Eq.~(\ref{Vmain}) and omit the high-order correction terms.

In the optimization, the low-energy coefficients are obtained from a
$\chi^2$ fit to realistic scattering data (represented here by phase
shifts from the CD-Bonn potential~\cite{machleidt2001}) and deuteron
properties. The fitting procedure is identical to the one described in
Section~\ref{optimize}.
Figure~\ref{CDBonn_phaseshifts} presents the resulting phase shifts for a
selection of scattering channels.
%
%
\begin{figure}[p]
\includegraphics[scale=0.58]{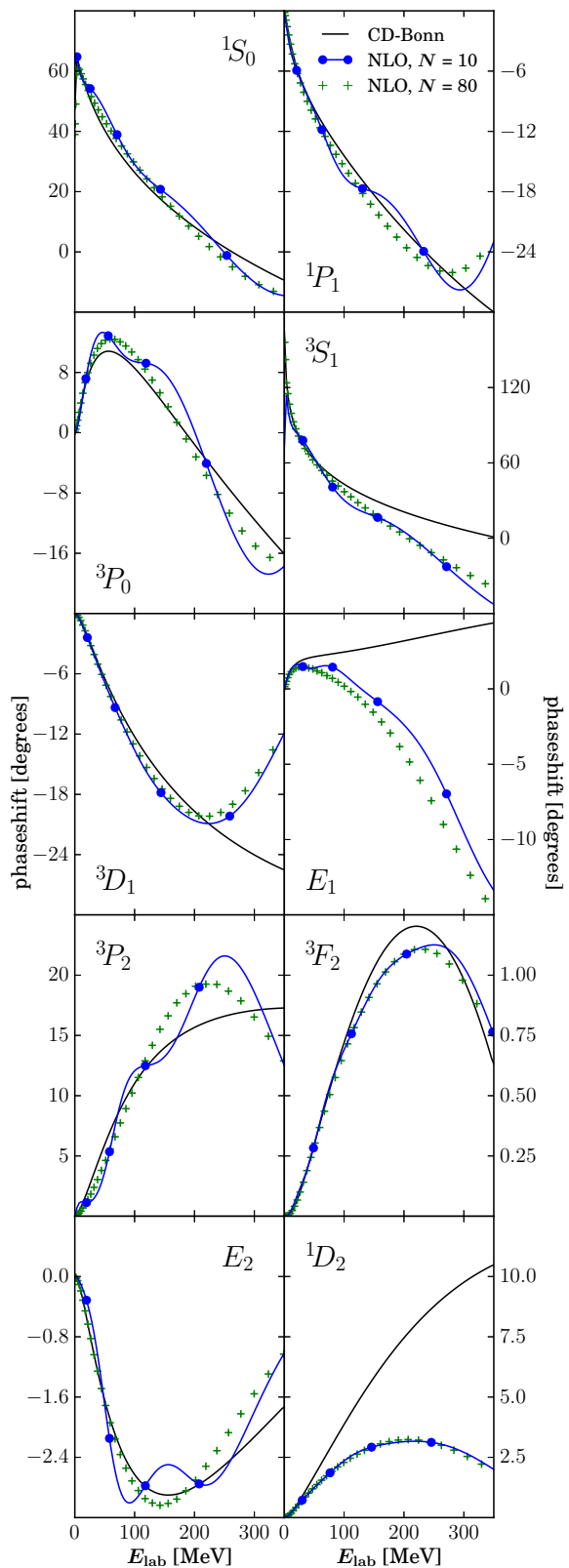}
\caption{(Color online) Phase shift from oscillator EFT at NLO 
with \mbox{$N_{\rm max} = 10$} (full
  blue circles and blue line) compared to those of the CD Bonn
  interaction (full black line) for partial wave channels as
  indicated. The green crosses show results from a large model space
  with $N_{\rm max}=80$.}
\label{CDBonn_phaseshifts}
\end{figure}
%
%
We reproduce the phase shifts over a large energy range for several
partial waves.  The ${^1\!D_2}$ channel is an obvious exception, as it
deviates more clearly from the CD-Bonn phase shifts at higher
energies, but we note that at NLO there is no LEC to adjust in this
channel.
For $E_1$, the deviations are also considerable, however, it is of NLO
quality, see Fig.~\ref{NLOsim_phaseshifts}.
For the deuteron, we obtain 
a good reproduction of the binding energy and radius,
and a reasonably well result for the quadrupole moment,
as shown in Table~\ref{CDBonnTable}.
The LECs, also shown in  Table~\ref{CDBonnTable}, are natural 
in size and similar to NLO$_{\rm sim}$;
the largest deviation is for the LEC $C_{^1\!P_1}$ which is
only about $\nicefrac{1}{3}$ of its NLO$_{\rm sim}$ counterpart.
%
%
\begin{table}[bt]
\begin{tabular}{| l d d d | }
\hline \hline
        & \multicolumn{1}{c}{\hspace{10pt} $N_{\rm max}=10$} & \multicolumn{1}{c}{\hspace{10pt} $N_{\rm max}=80$}   &  \multicolumn{1}{c|}{\hspace{10pt} experiment}   \\
\hline
$E_d \ [\text{MeV}]$ &-2.227 & 2.221 & -2.225\\
$r_d \ \ [\text{fm}]$     &1.984 & 1.961 &1.976\\
$Q_d\ [\text{fm}^2]$ & 0.229& 0.259&0.286\\
$P_d$                      & 0.026& 0.028 & \\
\hline
$\tilde{C}^{(np)}_{^1\!S_0}$ & -0.140992  & -0.139687 & \\
$\tilde{C}^{(pp)}_{^1\!S_0}$ & -0.13991    & -0.13855   & \\
$\tilde{C}^{(nn)}_{^1\!S_0}$ & -0.14046    & -0.13913  & \\
$C_{^1\!S_0}$                      & \hphantom{-}1.2470 & \hphantom{-}1.2601 & \\
$\tilde{C}_{^3\!S_1}$            &  -0.1604 & -0.1825 & \\
$C_{^3\!S_1}$                      & -0.682   & -0.361 & \\
$C_{E_1}$                           &  \hphantom{-}0.250 & \hphantom{-} 0.236 & \\
$C_{^3\!P_0}$                      &  \hphantom{-}1.1629 & \hphantom{-}1.1896 & \\
$C_{^1\!P_1}$                      &  \hphantom{-}0.348   & \hphantom{-}0.366 & \\
$C_{^3\!P_1}$                      &  -0.336 & -0.320 & \\
$C_{^3\!P_2}$                      & -0.1647& -0.1697& \\
 \hline \hline
\end{tabular}
\caption{Deuteron properties (as in Table~\ref{NLOsimTable}) and LECs
  from effective interactions in \mbox{$N_{\rm max} = 10$} and 80
  shells obtained from a fit to CD-Bonn phase shifts and deuteron
  properties.  The oscillator frequencies are $\hbar\omega$ = 24 MeV
  and \mbox{4 MeV} for \mbox{$N_{\rm max}$ = 10} and 80, respectively.  }
\label{CDBonnTable}
\end{table}
%
%

We utilize the NLO interaction in coupled-cluster calculations at the
singles and doubles level (CCSD)~\cite{kuemmel1978,
  shavittbartlett2009, hagen2014} of the nuclei $^4$He, $^{16}$O,
$^{40}$Ca, $^{90}$Zr, and $^{132}$Sn.  The coupled-cluster
calculations use Hartree-Fock bases that are unitarily equivalent to
the oscillator bases but lead to improved results.  In these
calculations, we keep the oscillator spacing fixed at
$\hbar\omega=24$~MeV in the spirit of oscillator EFT.  We employ model
spaces from $N^{\rm lab}=10$ up to $N^{\rm lab}=16$.  In the
oscillator basis, the potential is always restricted to $N_{\rm
  max}=10$, while the kinetic energy is used in the entire model
space.  The results are shown in Table~\ref{vAB_data}.  The
\mbox{$N^{\rm lab}=16$} point is used to gauge convergence of the
results.  For the light nuclei, energies and radii are practically IR
converged already for \mbox{$N^{\rm lab} = 10$} because $r^2 \ll L^2$.
\begin{table}[tb]
\begin{tabular}{|c|ccccc|}
\hline \hline
$N^{\rm lab}$& $^{4}\rm{He}$ & $^{16}\rm{O} $& $^{40}\rm{Ca}$ & $^{90}\rm{Zr}$ & 
 $^{132}\rm{Sn}$ \\
\hline
&  \multicolumn{5}{c|}{$E_{\rm CCSD}$ [MeV]} \\
\hline
 10 & -31.57 & -142.89 & -402.0 & -918.4 & -1230.0 \\
 12 & -31.57 & -142.92 & -402.4 & -923.1 & -1249.3 \\
 14 & -31.57 & -142.93 & -402.5 & -924.6 & -1255.6 \\ 
 16 & -31.57 & -142.93 & -402.5 & -925.1 & -1258.3 \\
$\infty$&-31.57& -142.93 & -402.5 & -925.4 & -1260.1 \\
\hline
exp & -28.30 & -127.62 & -342.1 & -783.9 & -1102.9 \\
\hline
\hline
&  \multicolumn{5}{c|}{$r^2 \ \rm [fm^2]$} \\
\hline
10 & 1.78 & 4.14 & 6.58 & 9.70 & 11.60 \\
12 & 1.78 & 4.15 & 6.60 & 9.77 & 11.79 \\
14 & 1.78 & 4.15 & 6.60 & 9.80 & 11.85 \\
16 & 1.78 & 4.15 & 6.61 & 9.82 & 11.88 \\
$\infty$ &  1.78 & 4.15 & 6.61 & 9.83 & 11.89\\
\hline
exp & 2.12 & 6.60 & 11.41 & 17.57 & 21.57 \\
\hline \hline
\end{tabular}
\caption{ Ground-state energies $E_{\rm CCSD}$ and squared
  point-proton radii $r^2$ for nuclei ranging from $^4$He to
  $^{132}$Sn from CCSD calculations in many-body model spaces built
  from single-particle oscillator bases with $N^{\rm lab} = 10, \dots,
  16$.
In the relative frame the oscillator EFT interaction is defined in the
$N_{\rm max} = 10$ model space, and the Hamiltonian matrix elements
outside this space are comprised of the kinetic energy only.
Experimental data taken from~\cite{wang2012,angeli2013}.
The experimental point-proton radii are extracted from charge radii
by correcting for the finite sizes of the proton and neutron, and the
Darwin-Foldy term.
}
\label{vAB_data}
\end{table}

 The convergence with respect to $N^{\rm lab}$ is
very fast compared to other interactions from chiral EFT with a
cutoff of around $\Lambda_\chi\approx450$~MeV~\cite{roth2014}.
For IR extrapolations of ground-state energies $E$ we employ~\cite{furnstahl2012}
 \be
\label{Eextra}
E(L)=E_\infty + a e^{-2k_\infty L} ,
\ee
where  $L\approx\sqrt{2(N^{\rm lab}+\nicefrac{7}{2})}b$ is the IR length~\cite{more2013}.
In the fit, we employ a theoretical uncertainty
\mbox{$\sigma\equiv\exp{(-2 k_\infty L)/(k_\infty L)}$} to account for
omitted corrections beyond the leading-order result~(\ref{Eextra}).  
The difference between the IR extrapolated energies and the energy at
$N^{\rm lab}=16$ is much smaller than both, the uncertainty of about
7\% in correlation energy due to the CCSD approximation, and the
uncertainty expected from higher orders in the chiral expansion.

Figure~\ref{Sn132_erg_rad} (top) shows the result of an IR
extrapolation for the ground-state energy of $^{132}$Sn. Here, we
plotted energies in $N^{\rm lab}=10$ to 16 as a function of the IR
length $L$ and also show the exponential IR
extrapolation~\cite{furnstahl2012}
The
inset of Fig.~\ref{Sn132_erg_rad} confirms the exponential convergence of the energy.
\begin{figure}[bt]
\includegraphics[scale=0.51]{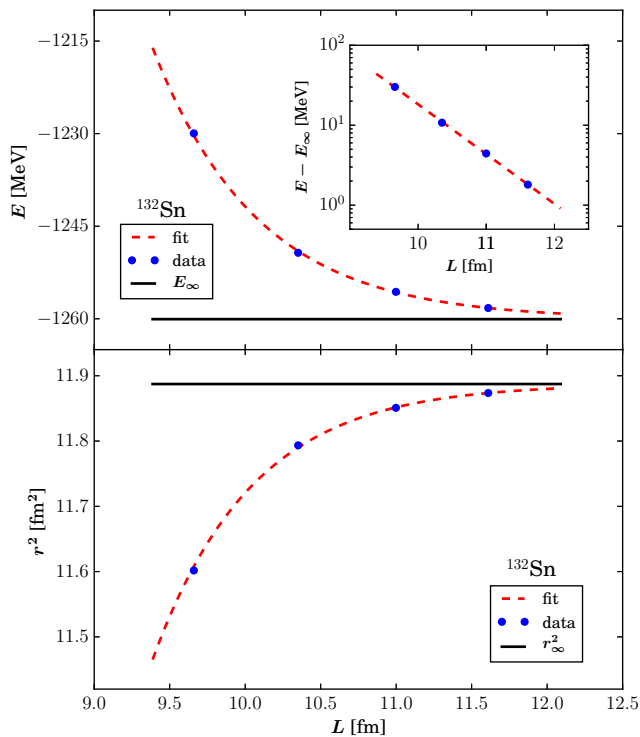} 
\caption{(Color online) IR convergence of the ground-state
  energy (top) and squared radius (bottom) of $^{132}$Sn computed
  from CCSD calculations using the oscillator EFT interaction.
   Data is shown as blue points, the 
  extrapolations as red dashed lines, and the asymptotic energy and radius as
  solid black lines. 
  The inset on the top shows that the ground-state energy is
  approached exponentially in the IR length $L$.
  }
\label{Sn132_erg_rad}
\end{figure}

Figure~\ref{Sn132_erg_rad} (bottom) shows the results of an IR
extrapolation for the ground-state expectation value of the squared
radius of $^{132}$Sn.  We plotted squared point-proton radii in model
spaces with $N^{\rm lab}=10$ to 16 as a function of the IR length $L$
and performed a $\chi^2$ fit to the IR
extrapolation~\cite{furnstahl2012}
\be
\label{rextra}
r^2(L) = r_\infty^2 - b(k_\infty L)^3 e^{-2k_\infty L} .
\ee
In the $\chi^2$ fit of the radius, we refit the constant $k_\infty$
and employed theoretical uncertainties \mbox{$\sigma\equiv k_\infty
L\exp{(-2 k_\infty L)}$} to account for corrections to the
leading-order result~(\ref{rextra}).
Numerical results for the radii are listed in Table~\ref{vAB_data}.
The results from IR extrapolations involving only the data points
from $N^{\rm lab} = 10$ to 14 yield $E_\infty = -1259.2$~MeV
and $r_\infty^2 = 11.88$~fm$^2$ for $^{132}$Sn, which are close to
the previous results including the \mbox{$N^{\rm lab} = 16$} points.

The oscillator EFT interaction at NLO overbinds nuclei by about
1~MeV per nucleon, and radii are too small.
We note that the difference between theoretical and experimental values
for the ground-state energies seem to be consistent with expectations
from an NLO interaction.
We also note that three-nucleon forces entering at 
next-to-next-to-leading order will be part of the saturation mechanism~\cite{hebeler2011, ekstrom2015}.

The rapid convergence of ground-state energies and radii in oscillator
EFT suggests that chiral cutoffs of $\Lambda_\chi\approx 450$~MeV are
reasonable in nuclear-structure computations of heavy nuclei. 
Similar to renormalization group transformations the oscillator EFT
reduces the mismatch between the tail of the momentum-space regulator
and the oscillator space.


\section{Summary}
\label{summary}

We developed an EFT directly in the oscillator basis. In this approach
UV convergence is implemented by construction, and IR convergence can
be achieved by enlarging the model space for the kinetic energy only.
We discussed practical aspects of the oscillator EFT and gave
analytical expressions for the efficient calculation of matrix
elements in oscillator EFT from their continuous-momentum
counterparts.  Within the $J$-matrix approach we computed phase shifts
while working exclusively in the oscillator basis. Our results suggest
that oscillations in the phase shifts appear when the UV cutoff
imposed by the oscillator basis cuts into the high-momentum tails of
the chiral interaction.  To validate the oscillator EFT approach we
reproduced a chiral interaction at NLO.  Finally, we developed a
chiral NLO interaction in oscillator EFT by optimizing the LECs to
CD-Bonn phase shifts and experimental deuteron data.  Coupled-cluster
calculations for nuclei from $^4$He up to $^{132}$Sn exhibit a rapid
convergence for ground-state energies and radii. These results suggest
that the oscillator EFT is a promising candidate to facilitate {\it ab
  initio} calculations of heavy atomic nuclei based on interactions
from chiral EFT using currently available many-body methods.


\begin{acknowledgments}
We are grateful to R.~J.~Furnstahl and S.~K\"onig for helpful discussions and comments on the manuscript. We also thank J.~Rotureau for helpful discussions.  This material is based
  upon work supported in part by the U.S.  Department of Energy,
  Office of Science, Office of Nuclear Physics, under Award Numbers
  DE-FG02-96ER40963 (University of Tennessee),
  DE-SC0008499 (SciDAC-3 NUCLEI Collaboration), the Field
  Work Proposal ERKBP57 at Oak Ridge National Laboratory (ORNL), and
  under contract number DEAC05-00OR22725 (ORNL).  S.B. gratefully
  acknowledges the financial support from the Alexander-von-Humboldt
  Foundation (Feodor-Lynen fellowship).
\end{acknowledgments}


\appendix*

\section{Matrix elements in oscillator EFT}
\label{appendix}

In this Appendix we give an alternative motivation for the
computation~(\ref{Vmain}) of matrix elements in oscillator EFT.  These
results are well known for DVRs, see, e.g.,
Ref.~\cite{littlejohn2002}.  The projection onto a finite oscillator
space is based on the usual scalar product
\be
\label{scapro}
\langle \tilde{\Psi}_f | \tilde{\Psi}_g\rangle = \int\limits_0^\infty {\rm d}k k^2 \tilde{\Psi}_f(k)\tilde{\Psi}_g(k)  
\ee
for radial wave functions in momentum space.  The projection operator
onto the finite oscillator space at fixed angular momentum $l$ is
\be
\label{projector_l}
\hat{\Pi}_l\equiv \sum_{\nu = 0}^N
|\phi_{\nu,l}\rangle\langle\phi_{\nu,l}| .  
\ee 

Let us alternatively consider a different projection operator, which
is based on a different scalar product. We write momentum-space
wave functions with angular momentum $l$ as
\be
\tilde{\Psi}_f(k) = \sqrt{2 b^3} (kb)^l e^{-{1\over 2}(kb)^2} f(k^2 b^2) .
\ee
Here, $f(x)$ is a polynomial in $x=k^2b^2$. It is clear that any
square-integrable wave function can be built from such polynomials
(with Laguerre polynomials $L_n^{l+\nicefrac{1}{2}}$ being an
  example). We define the (semi-definite) scalar product 
  $(\cdot | \cdot)$
  of two
  functions $\tilde{\Psi}_f$ and $\tilde{\Psi}_g$ as
\be
\label{scalar}
(\tilde{\Psi}_f | \tilde{\Psi}_g) \equiv \sum_{\mu=0}^N w_{\mu,l} f(x_{\mu,l}) g(x_{\mu,l}) .
\ee
Here, the weights are from Eq.~(\ref{weight}) and the abscissas
$x_{\mu,l}\equiv k_{\mu,l}^2b^2$ are the zeros of the associated Laguerre
polynomial $L_{N+1}^{l+\nicefrac{1}{2}}$ as demanded by
$(N+1)$-point Gauss-Laguerre quadrature.  Equation~(\ref{scalar})
defines only a semi-definite inner product, because
$(\tilde{\Psi}_f|\tilde{\Psi}_f) = 0$ for any polynomial $f$ with
$f(x_{\mu,l})=0$ for $\mu=0,\dots,N$.

Several comments are in order. First, this semi-definite scalar product
is identical to the standard scalar product~(\ref{scapro}) 
for wave functions $\tilde{\Psi}_f(k)$ that are limited to
the finite oscillator space. To see this, we note that
\ba
\langle \tilde{\Psi}_f|\tilde{\Psi}_g\rangle &=& \int\limits_0^\infty {\rm d} k k^2
\tilde{\Psi}_f(k)\tilde{\Psi}_g(k)\nonumber\\
&=& \int\limits_0^\infty {\rm d} x x^{l+\nicefrac{1}{2}} e^{-x} f(x) g(x) \nonumber
\\
&=&
\sum_{\mu=0}^N w_{\mu,l} f(x_{\mu,l}) g(x_{\mu,l}) . 
\ea
Here, we introduced the dimensionless integration variable
$x\equiv k^2b^2$ in the second line and employ $(N+1)$-point Gauss-Laguerre
quadrature in the third line. We note that $(N+1)$-point
Gauss-Laguerre quadrature is exact for polynomials of degree up to and
including $2N+1$, i.e., it is exact for polynomials $f, g$ spanned by
\mbox{$L_n^{l+\nicefrac{1}{2}}(x)$} with $n=0,\ldots, N$.

This implies that the basis functions $\tilde{\psi}_{n,l}(k)$ with
\mbox{$n=0,\ldots,N$} in the finite oscillator space remain a basis under the
semi-definite scalar product, i.e.
\be
(\tilde{\psi}_{n,l}|\tilde{\psi}_{n',l}) = \delta_n^{n'} .
\ee
Second, we note that both scalar products also agree for \mbox{$N\to\infty$} because
Gauss-Laguerre integration becomes exact in this limit. 

Rewriting the weights $w_{\nu,l}$ in Eq.~(\ref{weight}) as
\ba
\label{w2}
w_{\nu,l} &=& \frac{2 b^3 (k_{\nu,l} b)^2 (k_{\nu,l} b)^{2l} e^{-(k^2_{\nu,l} b^2)}}{(N+1)(N+l+\nicefrac{3}{2}) \left[\tilde{\psi}_{N,l}(k_{\nu,l})\right]^2}\nonumber\\
&=&2 b^3 (k_{\nu,l} b)^l e^{-{1\over 2}(k^2_{\nu,l} b^2)} c_{\nu, l}^2 ,  
\ea 
and using Eq.~(\ref{norm}) yields another useful expression for the
scalar product~(\ref{scalar})
\ba
\label{sca1}
(\tilde{\Psi}_f | \tilde{\Psi}_g) 
&=& \sum_{\mu=0}^N c_{\mu, l}^2 \tilde{\Psi}_f(k_{\mu,l})\tilde{\Psi}_g(k_{\mu,l}) .
\ea

A main difference between the scalar products of Eq.~(\ref{scapro})
and Eq.~(\ref{sca1}) arises when one compares the wave function
\be
\langle k, l|\phi_{\nu, l}\rangle = \tilde{\phi}_{\nu,l}(k) 
\ee
of Eq.~(\ref{overgen}) with the corresponding scalar product
\ba
\label{kphi}
(k,l|\tilde{\phi}_{\nu,l}) &=& \sum_{\mu=0}^N
c_{\mu, l}^2 {\delta(k-k_{\mu,l})\over k^2}\tilde{\phi}_{\nu, l}(k_{\nu,l}) \nonumber\\
&=& 
c_{\nu, l} {\delta(k-k_{\nu,l})\over k_{\nu,l}^2} . 
\ea
Here, we used Eq.~(\ref{sca1}) and Eq.~(\ref{overgen}) implying
\mbox{$\tilde{\phi}_{\mu,l}(k_{\lambda,l})=c_{\lambda,l}^{-1} \delta_\mu^\lambda$}. 
Clearly, the Fourier-Bessel transform~(\ref{overgen}) of the discrete
momentum eigenstate $\phi_{\nu,l}$ has a complicated momentum
dependence, while $(k,l|\tilde{\phi}_{\nu,l})$ is simply a rescaled
$\delta$ function. This simple view is consistent with
naive expectation of a momentum eigenstate. 

We are now in the position to compute matrix elements based on the
inner product~(\ref{scalar}). We note that
\ba
\label{ident}
\hat{V}|\phi_{\mu, l}) &=& 
\hat{V}\int\limits_0^\infty {\rm d}k k^2 |k,l\rangle (k,l|\phi_{\mu,l}) \nonumber\\
&=& \hat{V}|k_{\mu,l}\rangle c_{\mu,l}  .
\ea
Here, we used Eq.~(\ref{kphi}). Repeating the procedure on the bra side yields
\ba
(\phi_{\nu,l'}|\hat{V}|\phi_{\mu, l}) = \langle k_{\nu,l'}, l'|\hat{V}|k_{\mu,l}, l\rangle c_{\nu,l'} c_{\mu,l} , 
\ea
and this is Eq.~(\ref{Vmain}). We repeat that this derivation of
  the interaction matrix element is based on the scalar
  product~(\ref{sca1}) and not on the usual scalar
  product~(\ref{scapro}) for square integrable functions. We
  argue that the former scalar product is more natural considering the
   discrete momentum mesh that is employed in oscillator
  EFT.

For another view on the semi-definite scalar product we consider
the projection operator
\be
\label{proj_new}
\hat{P}_l \equiv \sum_{n=0}^N |\tilde{\psi}_{n,l})(\tilde{\psi}_{n,l}| . 
\ee
Indeed, $\hat{P}_l^2=\hat{P}_l$. This
makes it interesting to consider the projected wave function
\ba
\label{projeq}
\left(\hat{P}_l\tilde{\Psi}_f\right)(k_{\mu,l}) &=& \sum_{n=0}^N (\tilde{\psi}_{n,l}|\tilde{\Psi}_f) \tilde{\psi}_{n,l}(k_{\mu,l})\nonumber\\
&=& \sum_{\nu=0}^N c_{\nu, l}^2 \tilde{\Psi}_f(k_{\nu,l})\sum_{n=0}^N \tilde{\psi}_{n,l}(k_{\nu,l})\tilde{\psi}_{n,l}(k_{\mu,l})\nonumber\\
&=& \tilde{\Psi}_f(k_{\mu,l}) .
\ea
Here, we evaluated the sum over $n$ using Eq.~(\ref{useful}) when
going from the second to the third line. 
Equation~(\ref{projeq}) shows that the projected wave function agrees
with the full wave function at the discrete momenta $k_{\mu,l}$. In other
words, the projection based on the semi-definite scalar product yields
wave functions in finite oscillator spaces that agree with the
unprojected wave functions at the physical momenta. 

Let us give another interpretation of the projection
$\hat{P}_l$. One finds for the scalar product~(\ref{sca1}) 
\ba
\label{sca2}
\lefteqn{
(\tilde{\Psi}_f | \tilde{\Psi}_g) = \sum_{\mu=0}^N c_{\mu, l}^2 \tilde{\Psi}_f(k_{\mu,l})\tilde{\Psi}_g(k_{\mu,l})} \nonumber \\
&=& \langle\tilde{\Psi}_f| \left(\sum_{\mu =0}^N  |k_{\mu,l},l\rangle c_{\mu, l}^2 \langle k_{\mu,l}, l| \right) |\tilde{\Psi}_g\rangle .
\ea
Thus, the scalar product $(\tilde{\Psi}_f | \tilde{\Psi}_g)$ can be
viewed as a matrix element of the  operator in the brackets.


%

\end{document}